\documentclass[a4paper,10pt]{article}

\usepackage{graphicx}
\usepackage{epsfig}
\usepackage{amssymb}
\usepackage{amsmath}

\title{Quantum critical point in the spin glass-antiferromagnetism competition
for fermionic Ising Models}
\author{F. M. Zimmer and S. G. Magalh\~aes\footnote{E-mail address: ggarcia@ccne.ufsm.br}\\
Departamento de F\'\i sica, Universidade Federal de Santa Maria 
\\ 97111-900 Santa Maria, RS, Brazil.}
\date{}
\begin{document}
\maketitle

\begin{abstract}
The competition between spin glass ($SG$) and antiferromagnetic order ($AF$) is
analyzed in two sublattice fermionic Ising models in the presence of a transverse $\Gamma$ and a parallel $H$ magnetic fields.
The exchange interaction follows a Gaussian probability distribution with mean $-4J_0/N$ 
and standard deviation $J\sqrt{32/N}$, but only spins in different sublattices can interact. 
The problem is formulated in a path integral formalism, where the spin operators have been expressed as 
bilinear combinations of Grassmann fields. 
The results of two fermionic models are compared. In the first one, the diagonal $S^z$ operator has four 
states, where two eigenvalues vanish (4S model), which are suppressed by a restriction in the two states 2S model.
The replica symmetry ansatz and the static approximation have been used to obtain the free energy.
The results are showing in phase diagrams $T/J$ ($T$ is the temperature) {\it versus} $J_{0}/J$, $\Gamma/J$, 
and $H/J$. When $\Gamma$ is increased, $T_{f}$ (transition temperature to a nonergodic phase) reduces and the 
Neel temperature decreases towards a quantum critical point.
The field $H$ always destroys $AF$; however, within a certain range, it favors the frustration.
Therefore, the presence of both fields, $\Gamma$ and $H$, produces effects that are in competition. 
The critical temperatures are lower for the 4S model and it is less sensitive to the magnetic couplings 
than the 2S model.

\end{abstract}

%\begin{keyword}
%Quantum spin glass; Transverse field; Antiferromagnetism
%\PACS 64.60.Cn; 75.30.M; 05.50.+q
%\end{keyword}
%\end{frontmatter}

% main text

\section{Introduction}
\label{}
There are now several examples of competition between antiferromagnetism (AF) 
and spin glass (SG) in strongly correlated systems as, for instance, 
heavy fermions and high $T_{c}$ superconductors \cite{heavyfermions,highTc}. 
In some of these systems, 
there is also a new physics involved with the presence of a quantum critical point (QCP) and deviation of the 
Fermi liquid behavior, so called Non-Fermi liquid (NFL) behavior. 
This raises the necessity of obtaining a solid framework to describe 
those systems where the fermions are operative to give origin to
physical processes, 
such as 
the Kondo effect,     
in connection with the presence of AF and frustration driven by disorder,  
particularly, at low temperatures where quantum effects become important.

The complexities of such description can be estimated by the controversies 
\cite{hyun} involved in the quantum Ising SG which has been investigated by several techniques 
\cite{SG}. For instance,  one of the most interesting open issues is 
whether the quantum tunneling between the local degenerated minimum of the free energy 
is able to stabilize the replica symmetric solution of the problem.   
Recently,  
the fermionic Ising SG 
in the presence of a transverse magnetic field 
$\Gamma$ has been studied  using the functional integral
formalism \cite{Alba2002}.
The spins have been represented by bilinear combinations 
of the fermion operators. This fermionic problem has been presented in two versions. In the first one, the 
fermionic spin operators have maintained their four natural eigenvalues, 
where two of them 
are non-magnetic. In the second one, it has been retained
only the 
magnetic ones due to an imposed constraint to garantee complete equivalence between the spin 
and the fermionic problem \cite{Wiethege}.
The free energy has been obtained in both versions 
within the replica symmetric theory and the static approximation. 
It has been found that 
the freezing  
temperature $T_{f}(\Gamma)$ decreases to a QCP when $\Gamma$ enhances in a second order 
type transition. 
In that formulation, 
the replica symmetric solution is unstable in the entire SG region.  
The fermionic formulation is also a natural tool to study the interplay between frustration driven by disorder and, for instance, 
Kondo effect at low temperature when quantum tunneling is important. Actually, quite recently, 
the SG problem with transverse field  has been studied successfully in the disordered Kondo lattice 
with this technique \cite{AlbaCoqblin}. 

The purpose of the present work is to investigate the SG/AF competition in the context of the fermionic representation 
for the Ising spins variables when a magnetic field is applied with two components: one parallel ($H$) and other 
transverse ($\Gamma$) to the $z$ direction. 
The component $\Gamma$ introduces a spin flipping mechanism in the problem  
which can lead the phase boundaries to a QCP \cite {Alba2002}. 
The model used here is the fermionic version 
of the Korenblit-Shender (KS) model \cite{korenblit} 
introduced to study SG/AF competition with classical Ising 
spins. In the KS model, there are two sublattices, but only spins in distinct 
sublattices are allowed to interact 
 with a random Gaussian coupling $J_{ij}$.  The presence of the magnetic field $H$ 
breaks the symmetry between the sublattices. 
As consequence, it  can introduce unusual effects as it has been shown 
in the mean field theory with replica symmetry for the classical interlattice frustrated model \cite{korenblit}.    
For instance, 
the field $H$ can favor frustration within a certain range.  
This is in contrast with the well known result from the Sherrington-Kirkpatrick  model (SK) \cite{SK} 
where the freezing 
temperature $T_{f}$, associated with the Almeida-Thouless instability,
decreases monotonically with the field $H$. 
In fact, the dependence of the random internal field with $H$ in the KS 
model could explain its odd behavior \cite{korenblit}. The internal field 
applied $h_{p}$ in a particular sublattice ($p=a$, $b$) depends on the magnetization $m_{p'}$ and  
the SG order parameter $q_{p'}$ ($p'\neq p$). Due to 
the break of the symmetry between the two sublattices, when $H$ is increased 
in a particular range, one may have a nonmonotonic behavior in the order parameters 
$m_{p'}$, $q_{p'}$ and, consequently, 
in the $h_{p}$.
Therefore, $T_{f}$ can be enhanced and the non-trivial ergodicity breaking region (SG) is enlarged.

This leads several issues for 
the SG/AF competition in the fermionic representation of 
the KS model, when both field components ($H$ and $\Gamma$) are applied. 
Is the scenario described previously for the classical KS model preserved or not 
even for $\Gamma=0$ case? The answer is not obvious because the fermionic representation of the 
spin operators introduces 
an important difference as compared with 
its classical counterpart. The replica diagonal spin glass order parameter  
for the two sublattices is not constrained to the unity. It must be solved 
together with the other order parameters, the 
replica off-diagonal SG order parameter and the magnetization for the 
two sublattices. As consequence, it would introduce a new component in the random internal field. 
If the previous question is answered positively, 
is there any range of $H$ which can favor 
frustration when $\Gamma\neq 0$  as  in the classical KS? 
The presence of $H$,  
by a mechanism similar to the its classical counterpart, 
could enhance $T_{f}$   
while 
the $\Gamma$ component tends 
to suppress frustration leading the $T_{f}$ 
to a QCP \cite{Alba2002}. Therefore, the fields $H$ and $\Gamma$ enforce two 
competing mechanisms in the problem which could deeply affect the
SG/AF phase boundaries.

%It should be noticed that in the IF formalism
%\cite{Alba2002}, the SG phase boundary coincides with $T_{f}$ which is the key 
%physical quantity which caan answer the second question.}

The quantum mechanical partition function of the problem has been obtained 
following the approach 
introduced in reference \cite{Alba2002}. Therefore, the functional integral 
approximation is used to deal with 
non-commutativity of the spin operators which are represented 
by bilinear combinations of Grassmann 
variables.  
One important aspect of the fermionic representation 
of spins $S^{z}_{i}$ is that (see Ref. \cite{Alba2002}) 
it has four eingenvalues per site, where two of them are non-magnetic. 
For the rest of the paper, this representation is named  
4S model.  To recover the usual spin representation, a 
constraint is introduced in order to maintain only the magnetic eigenvalues. This version 
of the problem is named 
2S model. The disorder in the problem is  treated using the replica trick 
where the order parameters are obtained within the replica symmetric ansatz 
\cite{SK}.  
It should be noticed that the possible occupation of non-magnetic states  can produce 
differences concerning the phase boundaries between the two models. 
In fact, for the one lattice 
fermionic SG with transverse field \cite{Alba2002}, $(T_{f})_{2S}>(T_{f})_{4S}$ for 
$\Gamma<<\Gamma_{c}$ where $\Gamma_{c}$ is the value of the field at the QCP.
This also arises the question  how  these 
representations 2S and 4S respond (in terms of the phase boundaries) 
when $H$ and $\Gamma$ are turned on. 
In order to allow us a better comparision between the two representations of the problem, 
the 4S model is kept in the half-filling.

In the present approach, the time dependence of the spin-spin correlation functions 
has not been considered (static approximation) \cite{Bray}.  The main argument to justify 
the use of this approximation (see Ref. \cite{Alba2002,Magal04}), even at low temperature, 
is that our goal is to obtain the phase boundaries in the SG/AF competition in the 
spin fermionic representation and how these 
boundaries are affected by the presence of $H$ and $\Gamma$.  
As it can be seen elsewhere 
\cite{AlbaCoqblin,Magal042}, that approximation is quite reasonable if the intention is not 
to reveal the complex nature of phases at very low temperature, but mainly to 
produce phase boundaries which can 
mimic the experimental ones. 

It is hoped that the results presented here can be a first step to provide a 
framework which can allow us to study  the phase boundaries SG/AF competition 
present in several problems of many interacting fermions, such as the heavy fermion system 
$Ce_{2} Au_{1-x} Co_{x} Si_{3}$ \cite{Majundar} 
where disorder, competing RKKY interaction and 
the Kondo effect have produced a SG alike state and an AF phase where the Neel temperature 
seems to decrease towards a QCP with no trace of NFL behavior. %\cite{Magal042}.  

This paper has the following structure. In section II, the model is introduced,  the saddle point  
free energy and the corresponding order parameters are obtained. In section III, the numerical solutions 
of the order parameters allow  us to construct phase diagrams to show the AF/SG competition.
Particularly interesting, for the purposes 
of the present work, are the phase diagrams in the space temperature versus the 
components of the magnetic fields $H$ and $\Gamma$. It is also shown 
the behavior of the susceptibility. In the last section, the discussion 
of the results previously shown and the final remarks are presented.

\section{Model}
The model considered here is a fermionic Ising model \cite{AlbaMercedes} represented in two sublattices $a$ and $b$ 
where there are two 
magnetic fields applied: $\Gamma$ and $H$ transverse and parallel to the Ising  spins, respectively. One important point is that  only  
spins located in distinct sublattices can interact like in the KS model \cite{korenblit}. Thus, 
\begin{equation}
{\hat{ H}}= -\sum_{i_{a}j_{b}} J_{i_{a}j_{b}}\hat{S}_{i_{a}}^{z} \hat{S}_{j_{b}}^{z}
-2\sum_{i_a} \left(\Gamma\hat{S}_{i_{a}}^{x}+H \hat{S}_{i_{a}}^{z} \right)
-2\sum_{j_b} \left(\Gamma\hat{S}_{j_{b}}^{x}+H \hat{S}_{j_{b}}^{z} \right)
\label{ham}
\end{equation}
\noindent
where the sums are run over the $N$ sites of each sublattice ($a$ or $b$).
The coupling $J_{i_a j_b}$ is assumed to be a random variable with a Gaussian
distribution given by: 
\begin{equation}
P(J_{i_{a}j_{b}})=\sqrt{\frac{N}{64\pi J^2}}\exp\left[-\frac{\left(J_{i_{a}j_{b}}+4J_0/N\right)^2}{64 J^2}N\right]
\label{gaussian}~.
\end{equation}
The spin operators in Eq. (\ref{ham}) are defined as \cite{Alba2002}:
\begin{equation}
\hat S_{i_p}^{z}=\frac{1}{2}[\hat{n}_{i_p\uparrow}-\hat{n}_{i_p\downarrow}]~,
~~~~~~
\hat S_{i_p}^{x}=\frac{1}{2}[c_{i_p\uparrow}^{\dagger}c_{i_p\downarrow}
+c_{i_p\downarrow}^{\dagger}c_{i_p\uparrow}]
\label{ope}
\end{equation}
\noindent
where $\hat{n}_{i_p\sigma}=c_{i_p\sigma}^{\dagger}c_{i_p\sigma}$ is the 
number operator, $c_{i_p\sigma}^{\dagger}~(c_{i_p\sigma})$ are fermions 
creation (destruction) operators, with $\sigma=\uparrow$ or $\downarrow$ 
indicating the spin projections, and the sub-index $p=a$ or $b$ represents the
sublattice.   

In Eq. (\ref{ope}), the spins have been written as bilinear combination % in terms 
of fermion operators which act on
a space with four states per site ($\left|0~0\right\rangle$, $\left|\uparrow0\right\rangle$,  $\left|0\downarrow\right\rangle$, 
$\left|\uparrow\downarrow\right\rangle$). Therefore, $\hat{S}_{i_p}^{z}$ has four eigenvalues: 
$\pm 1/2$ (when there is one fermion in the
site $i_p$ ($\sigma=\uparrow$ or $\downarrow$)) and two  when
the site $i_p$ is unoccupied or double occupied ($\sigma=\uparrow$ and
$\downarrow$).  
In the present work, two formulations are considered: one unrestrained
the number of states of the $S_{i_p}^{z}$ (4S model), but it considers the average occupation of one fermion 
per site, another restrained the
occupation number to $\hat{n}_{i_p\uparrow}+\hat{n}_{i_p\downarrow}=1$ (2S model). This last representation 
allow us to study the problem avoiding the presence of unoccupied and double occupied states \cite{Alba2002}.

The partition function is given in the Lagrangian path 
integral formalism  where the spin operators are represented as anticommuting Grassmann fields ($\phi,~\phi^*$).
The partition function for the 2S model must consider only states that have one 
fermion per site. This restriction is obtained  by using the Kronecker $\delta$ 
function ($\delta(\hat{n}_{i_{p}\uparrow}+ \hat{n}_{i_{p}\downarrow}-1)$ $= 
\frac{1}{2\pi} \int_{0}^{2\pi} dx_{i_{p}}$ $e^{ix_{i_p}(\hat{n}_{i_{p}\uparrow}+
\hat{n}_{i_{p}\downarrow}-1)}$). Therefore, the partition function   
for both models can be represented in a compact form as \cite{Alba2002}:
\begin{equation}
Z\{\mu\}=\prod_{p=a,b}\prod_{i_{p}}\frac{1}{2\pi}
\displaystyle \int_{0}^{2\pi} dx_{i_{p}}e^{-\mu_{i_{p}}}
\int D(\phi^{*}\phi)~\exp \left( A\{\mu\}\right)
\label{eqz}
\end{equation} 
where $\mu_{i_p}=0$ for the  4S model,  which corresponds to the half-filling situation,
 or $\mu_{i_p}=ix_{i_p}$ for the
2S model. 
The  action $A\{\mu\}$ can be Fourier transformed in time. Thus, we have: 
\begin{equation}
A\{\mu\}=A_M^a+A_M^b+A_{SG}
\label{acao}
\end{equation}
with
\begin{equation}
A_{M}^p=\sum_{i_{p}}\sum_{\omega}\underline{\phi}_{i_{p}}^{\dag}(\omega)
\left[ i\omega+\mu_{i_{p}}
+\beta H \underline{\sigma}^{z} +\beta \Gamma \underline{\sigma}^{x} 
\right]\underline{\phi}_{i_{p}}(\omega)
\label{eqao},
\end{equation}
\begin{equation}
A_{_{SG}}=\sum_{\Omega}\sum_{i_{a}j_{b}}\beta J_{i_{a}j_{b}}S_{i_{a}}(\Omega)S_{j_{b}}(-\Omega)
\label{asg},
\end{equation}
\begin{equation}
S_{i_{p}}(\Omega)=\sum_{\omega}\underline{\phi}_{i_{p}}^{\dag}(\omega+\Omega)
\underline{\sigma}^z \underline{\phi}_{i_{p}}(\omega)
\label{eqs},
\end{equation}
\noindent
$\beta$ the inverse temperature and the matrices in Eqs. (\ref{eqao}, \ref{eqs}) are defined as:
\begin{equation}
\underline{\phi}_{i_p}(\omega)=\left[\begin{tabular}{c}$\phi_{i_p\uparrow}(\omega)$ \\ 
$ \phi_{i_p\downarrow}(\omega)$ \end{tabular}
\right]; ~
\underline{\sigma}^x=\left(
\begin{tabular}{cc}
$0$ & $1$\\  $1$ & $0$
\end{tabular}
\right); 
\ \
\underline{\sigma}^z=\left(
\begin{tabular}{cc}
$1$ & $0$\\  $0$ & $-1$
\end{tabular}
\right);
\ \
\label{pauli}
\end{equation}
\noindent
with the Matsubara's frequencies $\omega=(2m+1)\pi$ and $\Omega=2m\pi~ (m=0,\pm
1,\cdots)$. In this work,
the problem is analyzed in the static approximation, which 
considers only the term  when $\Omega=0$ in Eq. (\ref{asg}) \cite{Alba2002,Bray,AlbaMercedes}.

The free energy per site is obtained by using the replica method:
$\beta F =-\ln{Z}=-\lim_{n\rightarrow 0}$ $1/(nN)({{Z}}(n)-1)$, where $Z(n)\equiv\langle 
{{Z}}^n \rangle_{J_{i_{a}j_{b}}}$ is the configurational averaged replicated 
partition function. 
The average over $P(J_{i_{a}j_{b}})$ can be performed using the Gaussian distribution given in Eq. (\ref{gaussian}).Thus:
\begin{equation}
\begin{split}
{ Z}(n)&=\prod_{p=a,b}\prod_{i_p}\prod_{\alpha=1}^{n}
\frac{1}{2\pi}\int_{0}^{2\pi}dx_{i_p}^{\alpha}e^{-\mu_{i_p}^{\alpha}}
\int  D(\phi_\alpha^*,\phi_\alpha)
\exp\left\{ \sum_{\alpha=1}^{n}(A_{M}^{a,\alpha}\right.
\\
&\left.
+A_{M}^{b,\alpha})+\sum_{i_a j_b}\left[ \frac{16 \beta^2 J^2}{N}\left( \sum_{\alpha=1}^{n}
S_{i_a}^{\alpha}S_{j_b}^{\alpha}\right)^2
-\frac{4 \beta J_0}{N}\sum_{\alpha=1}^{n}S_{i_a}^{\alpha}S_{j_b}^{\alpha} 
\right] \right\}
\label{zn1}
\end{split}
\end{equation}
where $\alpha$ denotes the replica index and $S_{i_p}^{\alpha}\equiv S_{i_p}^{\alpha}(0)$. 
Eq. (\ref{zn1}) can be rearranged  reviewing
the sums over 
different sublattices by square sums
over the same sublattice \cite{korenblit}. Then, these 
quadratic terms are linearized by using the Hubbard-Stratonovich transformation. 
This transformation inserts auxiliary fields $\{M_3^\alpha$, $M_p^\alpha$, $Q_{3}^{\alpha\gamma}$, 
and $Q_{p}^{\alpha\gamma} \}$ on the partition function, therefore:
\begin{equation}
\begin{split}
{ Z}(n)=&\int d{ U}
\exp\left\{-N\left[\frac{\beta J_0}{2} \sum_{\alpha} \left( 
(M_{3}^\alpha)^2 + \sum_{p=a,b}(M_{p}^\alpha)^2\right)
\right.\right.\\
&\left.\left.
+\frac{\beta^2 J^2}{2}\sum_{\alpha\gamma}\left(
(Q_{3}^{\alpha\gamma})^2 +\sum_{p=a,b}(Q_{p}^{\alpha\gamma})^2\right)
-\frac{1}{N}\ln\Theta_{\alpha \gamma}\{\mu\}\right]\right\}
\label{zn2}
\end{split}
\end{equation}
\noindent where 
$\int d{ U}=\int_{-\infty}^{\infty}\prod_{r=a,b,3}
\prod_{\alpha}dM_{r}^{\alpha}
\int_{-\infty}^{\infty}\prod_{r=a,b,3}\prod_{\alpha\gamma}
dQ_{r}^{\alpha\gamma}$, and the functional part is expressed as: 
\begin{equation}
\begin{split}
\Theta_{\alpha\gamma}\{\mu\}&=\prod_{p=a,b}\prod_{i_p,\alpha}
\frac{1}{2\pi}\int_{0}^{2\pi}dx_{i_p}^{\alpha}e^{-\mu_{i_p}^{\alpha}}
\int  D(\phi_\alpha^*,\phi_\alpha)
\\
&\exp\left\{\sum_{p=a,b}\left[\sum_{\alpha}A_{M}^{p,\alpha}+
2\beta J_0\sum_{\alpha,i_p}\left(iM_{3}^{\alpha} + M_{p}^{\alpha}\right)
S_{i_p}^{\alpha}
\right.\right.
\\ 
&\left.\left.
+4\beta^2 J^2 \sum_{\alpha\gamma,i_p}\left(
Q_{3}^{\alpha\gamma} + iQ_{p}^{\alpha\gamma}\right)S_{i_p}^{\alpha}S_{i_p}^{\gamma}
\right] \right\}~.
\label{lambda}
\end{split}
\end{equation}

In the thermodynamic limit, the set of integrals in $\int d{ U}$
can be performed exactly by the steepest descent method, where the auxiliary fields
are given by the saddle point solutions:
\begin{align}
M_{3}^{\alpha}&=\frac{i 2}{N}\langle \sum_{p=a,b}\sum_{i_p}S_{i_p}^{\alpha} %+ 
\rangle =2im_{3}^{\alpha}~;~~
%\\
Q_{3}^{\alpha\gamma}=\frac{4}{N}\langle \sum_{p=a,b}\sum_{i_p}S_{i_p}^{\alpha}S_{i_p}^{\gamma}
\rangle =2 q_{3}^{\alpha\gamma}~;
\\
M_{p}^{\alpha}&=
\frac{2}{N}\langle \sum_{i_p}S_{i_p}^{\alpha}\rangle =m_{p}^{\alpha}
~;~~%,~~p=a,~ b~;
Q_{p}^{\alpha\gamma}=\frac{i 4}{N}\langle \sum_{i_p}S_{i_p}^{\alpha}S_{i_p}^{\gamma}
\rangle =i q_{p}^{\alpha\gamma}; ~~p=a,~b
\label{parametros}~
\end{align}
\noindent
where $\langle\dots\rangle$ denotes the average taken with respect to Eq. (\ref{lambda}).
These saddle point equations can be used to rewrite Eq. (\ref{zn2}) as:
\begin{equation}
\begin{split}
{ Z}(n)=\int d{ U}
\exp\left\{-N\left[\beta^2 J^2\sum_{\alpha\gamma}q_{a}^{\alpha\gamma}q_{b}^{\alpha\gamma}
-\beta J_0\sum_{\alpha}m_{a}^{\alpha}m_{b}^{\alpha} - 
\right.\right.
\left.\left.
\sum_{p} \ln\Theta_{\alpha\gamma}^{p}\{\mu\}
\right]\right\},
\label{zn3}
\end{split}
\end{equation}
\noindent
where: 
\begin{equation}
\Theta_{\alpha\gamma}^{p}\{\mu\}=\prod_{\alpha}\frac{1}{2\pi}\int_{0}^{2\pi}
dx_{p}^{\alpha}e^{-\mu_{p}^{\alpha}}\int D[\phi_p^{*}\phi_p]\exp\left[H_{p}^{eff}\right]~,
\label{lambdaeff}
\end{equation}
\noindent
with
\begin{equation}
H_{p}^{eff}=\sum_{\alpha}\left[A_{M,p}^{\alpha} -2\beta J_0 m_{p'}^{\alpha} S_{p}^{\alpha}\right] + 
 4\beta^2 J^2\sum_{\alpha\gamma} q_{p'}^{\alpha\gamma}S_{p}^{\alpha}
S_{p}^{\gamma}
\label{heff},
\end{equation}
\noindent 
and $p=a~(p'=b)$ or $p=b~ (p'=a)$.

At this stage, it is assumed the replica symmetric ``ansatz'' that considers the set 
$q_{p}^{\alpha \gamma}= q_{p}$ for all $\alpha\neq\gamma$, 
$q_{p}^{\alpha \alpha}= q_p + \bar{\chi}_p$,
and $m_{p}^{\alpha}=m_p$ for all $\alpha$. 
The physical quantity $\bar{\chi}_{p}=\chi^{*}_{p}/\beta$ where $\chi^{*}_{p}$ is the magnetic susceptibility when $J_{0}=0$. 
The sums over $\alpha$ in Eq. ({\ref{heff}}) produce quadratic terms again, which 
can be linearized introducing new auxiliary fields in the expression (\ref{lambdaeff}),
therefore:  
\begin{equation}
\Theta_{\alpha\gamma}^{p}\{\mu\}=\int_{-\infty}^{\infty}Dz_p
\left[\int_{-\infty}^{\infty}D\xi_p \frac{1}{2\pi}\int_{0}^{2\pi}dx_{p} e^{-\mu_{p}}
\Lambda_{eff}^{p}\right]^n
\label{funcional}
\end{equation}
\noindent where $Dy_p=dy_p \exp{(-y_p^2/2)}/\sqrt{2\pi}$,
and
\begin{equation}
\Lambda_{eff}^{p}=\int D[\phi_p^{*}\phi_p]\exp\left[
\sum_{\omega}\underline{\phi}_{{p}}^{\dag}(\omega)\underline{G}_{p}^{-1}(\omega)
 \underline{\phi}_{{p}}(\omega)\right]
\label{funcional1}~.
\end{equation}
The matrix $\underline{G}_{p}^{-1}(\omega)$ is given by:
\begin{equation}
\underline{G}_{p}^{-1}(\omega)=
\left(\begin{tabular}{cc}
$i\omega+\mu_{p}+\beta h_{p}$ & $\beta \Gamma$\\  
$\beta \Gamma$ & $i\omega+\mu_{p}-\beta h_{p}$
\end{tabular}\right)\ 
\label{e33}  
\end{equation}
\noindent where  
\begin{equation}
h_{p}= (H- J_0 m_{p'} +  J
\sqrt{2q_{p'}}z_p+ J \sqrt{2 \bar{\chi}_{p'}}\xi_p)
\label{hp}
\end{equation}
with $p'\neq p$. 
Therefore, the internal 
field $h_{p}$ applied in a particular sublattice depends  entirely on the order parameters 
of the other sublattice \cite{korenblit}.

The functional integral over the Grassmann variables in Eq. (\ref{funcional1}) can be 
calculated (see Ref. \cite{Negele}), thus, $\ln\Lambda_{eff}^{p}=\sum_{\omega}
\ln(\det{\underline{G}^{-1}_{p}(\omega))}$. In the same equation, the 
sum over the Matsubara's frequencies can be performed like references \cite{Alba2002,AlbaMercedes}  to give 
the following expression:
\begin{equation}
\Lambda_{eff}^{p}=2e^{\mu_p}[\cosh{\mu_p}+\cosh{(\beta\sqrt{\Delta_p})}]~
\label{soma}
\end{equation}
where $\Delta_p=h_p^2+\Gamma^2$.

Now, the restriction condition over  the number of states can be used for both models. For the 
4S model,  
the average occupation per site is one. It is obtained putting $\mu_p=0$, therefore the integral over $x_p$ is equal 
to the unity. For the 2S model  ($\mu_p=i x_p$), which corresponds to the spin formulation 
where there are only two states per site ($|\uparrow,0\rangle$, $|0,\downarrow\rangle$), the integral over 
$x_p$ is equal to zero. This situation is equivalent to choose an imaginary and temperature-dependent
chemical potential in the Popov-Fedotov method \cite{popov}. Therefore, the main difference between 
both models is that in the 2S model the contribution of the non-magnetic local states is 
exactly canceled, while in the 4S model it is adopted the half-filling situation. These results 
are used in Eq. (\ref{zn3}) and the free energy can be written as:
\begin{equation}
\beta F_{s}=\beta^2 J^2 (q_a\bar{\chi}_b+q_b\bar{\chi}_a +\bar{\chi}_a\bar{\chi}_b) -
\beta J_{0} m_a m_b -\sum_{p=a,b}\int_{-\infty}^{\infty}Dz_p \ln \Theta_{s,p}-\ln 4
\label{freeenergy}
\end{equation}
\noindent
with
\begin{equation}
\Theta_{s,p}=\frac{s-2}{2}+\int_{-\infty}^{\infty}D\xi_p
\cosh(\beta\sqrt{\Delta_p})~,
\label{difference}
\end{equation}
\noindent
where $s$ represents the states number allowed in each model. The order parameters $q_p,~\bar{\chi}_p$ and $m_p$ are given by extreme condition 
of the free energy (\ref{freeenergy}):
\begin{align}
m_{p}&=\int_{-\infty}^{\infty}Dz_{p}\frac{\int_{-\infty}^{\infty}D\xi_p h_p\sinh(\beta\sqrt{\Delta_p})/\sqrt{\Delta_p}}{\Theta_{s,p}}\label{em} ~,
\\
q_{p}&=\int_{-\infty}^{\infty}Dz_{p}\left(\frac{\int_{-\infty}^{\infty}D\xi_p h_p\sinh(\beta\sqrt{\Delta_p})/\sqrt{\Delta_p}}{\Theta_{s,p}}\right)^2~\label{eq},
\\
\bar{\chi}_{p}&=\int_{-\infty}^{\infty}Dz_{p}\frac{\int_{-\infty}^{\infty}D\xi_p
[h_{p}^{2}\cosh(\beta\sqrt{\Delta_p})/\Delta_p+\Gamma^2\sinh(\beta\sqrt{\Delta_p})/
(\beta\Delta_{p}^{3/2})] }{ \Theta_{s,p}}-q_p.\label{ex}
\end{align} 

In particular, when $\Gamma=0$, the set of integrals over $\xi_p$ can be performed 
analytically. In this case, it is easy to see that the parameter $\bar{q}_{p}\equiv \frac{4}{N}\langle
\sum_{i_p} S_{i_p}^{\alpha}S_{i_p}^{\alpha}\rangle=q_p+\bar{\chi}_p=1$ (Eq. (27)) 
for the 2S model. However, in the 4S model, due to the presence of non-magnetic states, $\bar{q}_p$ 
depends on the temperature and the internal field $h_p$. Nevertheless, near $T=0$, $\bar{q}_p$ can be 
expressed by using a low-temperature expansion:
\begin{equation}
\bar{q}_p=1-\frac{(s-2)}{2} \frac{T}{J}\sqrt{\frac{\pi}{4q_{p^{'}}}}
\exp\left[ -\frac{(\varrho_{p})^2}{2} - 
\frac{J\mbox{e}^{-(\varrho_{p})^2/2}}{T\sqrt{q_{p^{'}}\pi}}
\right],
\end{equation}
where $\varrho_{p}=\frac{J_0 m_{p^{'}}}{J\sqrt{2q_{p^{'}}}}$.
This result suggests that the occupation of non-magnetic states are exponentially small at low $T$.

It is well known that the replica symmetric solution can be unstable at low temperature in the spin glass phase \cite{Alba2002}, 
where the Almeida-Thouless eigenvalue $\lambda_{AT}$ becomes negative. 
The equation for the $\lambda_{AT}$ following as \cite{korenblit}: 
\begin{equation}
\lambda_{AT}=1-2(\beta J)^4\prod_{p=a,b}\int_{-\infty}^{\infty}Dz_{p}
\left(\frac{I_{p}}{\Theta_{s,p}^{2}}\right)^{2}
\label{almeida}
\end{equation}  
where
\begin{equation}
\begin{split}
I_{p}(z)=\Theta_{s,p}\int_{-\infty}^{+\infty}D\xi_p
\left[\frac{h_{p}^{2}}{\Delta_{p}}\cosh(\beta\sqrt{\Delta_{p}})+\frac{\Gamma^{2}}{\beta\Delta_{p}^{3/2}}
\sinh(\beta\sqrt{\Delta_{p}})\right]  \\
-\left(\int_{-\infty}^{+\infty}D\xi_p\frac{h_{p}}{\sqrt{\Delta_{p}}}\sinh(\beta\sqrt{\Delta_{p}})
\right)^{2}.
\end{split}
\end{equation}

\section{Results}

\begin{figure}
\begin{center}
	\includegraphics[angle=270,width=10.5cm]{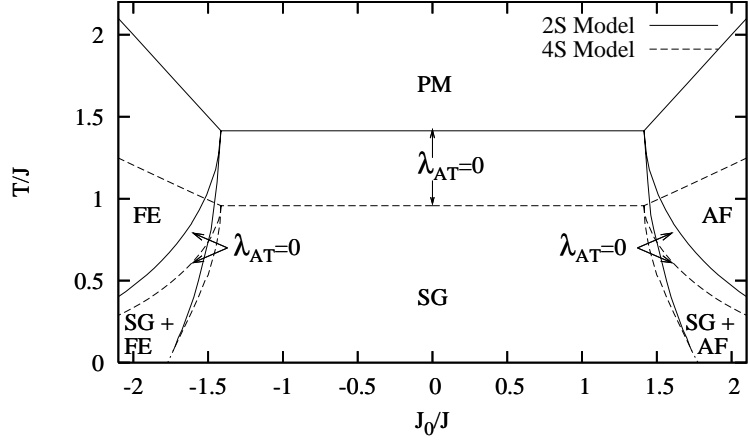}
\end{center}
\caption{Phase diagram $T/J$ {\it versus} $J_{0}/J$ for $H=\Gamma=0$. 
The solid lines correspond to the results obtained by the 
restricted model (2S), while the dashed lines correspond to the unrestricted model (4S).  The dotted 
lines are the extrapolation carried for lower temperature.
$\lambda_{AT}=0$ is the Almeida-Thouless line.}
\label{fig1}
\end{figure}
The phase diagrams showing the competition between SG and AF for both
models with two (2S) and four (4S) states  can be obtained solving 
the replica symmetric order parameters given by the set of equations (\ref{em})-(\ref{ex}). 
In the present work, the numerical solutions for $m_{p}$, $q_{p}$, $\overline{\chi}_{p}$ and 
$\lambda_{AT}$ (with $p=a$ or $b$) are studied 
by varying the three relevant parameters in the problem, which are $J_{0}/J$, $\Gamma/J$ and $H/J$.  
In this context, the AF phase is given by the order parameters $l_{m}=m_{a} - m_b\neq 0$ and 
$q_a-q_b\neq 0$; however, the same situation with the eigenvalue $\lambda_{AT}<0$ characterizes the 
mixed phase (SG+AF). It can be considered that, in the SG phase, the order parameters 
$l_m=0$ and $q_a=q_b\neq 0$ together with $\lambda_{AT}<0$.
In this quantum fermionic treatment, one can identify two situations for the diagonal 
component of the replica matrix ($\bar{q}_{p}=q_p+\bar{\chi}_p$).
For the 2S model with $\Gamma=0$, $\bar{q}_{p}=1$.
On the other hand, for the 4S model or for the 2S model when $\Gamma>0$, it is necessary to consider the 
coupling between diagonal and off-diagonal replica matrix elements. Consequently, 
the parameter $\bar{\chi}_p$ must be calculated simultaneously with $q_p$ and $m_p$, 
and hence it becomes relevant for determination of the remaining physical quantities.
In particular, for $H=0$, the Neel temperature $T_{N0}=T_{N}(J_{0},\Gamma,H=0)$ can be computed expanding 
the saddle point equations in powers of $m_p$. At the second order critical line $T_{N0}$, we can make $q_p=0$, 
therefore $\bar{\chi}=\bar{\chi}_{p}=1/(\beta_{c} J_{0})$ and 
\begin{equation}
m[1-\beta_c J_0(\frac{\frac{s-2}{2}+\int D\xi \xi^2 \cosh\beta_{c}\sqrt{\frac{2 J \xi^2}{J_0 \beta_{c}}+\Gamma^2}}{\frac{s-2}{2}+\int D\xi \cosh\beta_{c}\sqrt{\frac{2 J \xi^2}{J_0 \beta_{c}}+\Gamma^2}})]=0
\label{mc}~,
\end{equation}
where $\beta_c =1/T_{N0}$ and $m=m_p=-m_{p^{'}}$. For $\Gamma=0$, the Neel temperature is 
$T_{N0}=J_0/[\frac{s-2}{2}$ $\exp{(\frac{-J^2}{T_{N0} J_0})}+1]$, while for $T_{N0}$ close to zero, the critical 
value of the transverse field is given by analytical solution of Eq. (\ref{mc}): $\Gamma_N=J_0 + 2 J^2/J_0$, for both models.

In Fig. (1), it is shown the diagram $T/J$ ($T$ is the temperature) {\it versus} $J_{0}/J$ when $\Gamma$ and $H$ are zero. 
For that particular situation, the results for 
2S and 4S models are qualitative equivalent to the KS model with classical Ising spins.  Actually, for the 2S model, 
the Fischer relation \cite{hyun} is recovered $\overline{\chi}_{a,b}\equiv\chi_{a,b}^{*}/\beta=1-q_{a,b}$
($\chi_{a,b}^{*}$ is 
the linear susceptibility for $J_0$ equal to zero).
Therefore, the 2S model reproduces exactly  
the KS results \cite{korenblit}.    
For high temperature and high degree of frustration $\alpha\equiv\left[J_{0}/J\right]^{-1}$, 
a paramagnetic phase (PM) is found.
The AF solution for both models can be found 
in the region where the 
degree of frustration 
$\alpha$
is small enough. In this case,  
$m_{a}=-m_{b}$ \cite{korenblit}. For increasing $\alpha$ and decreasing temperature, the Almeida-Thouless line 
in Fig. (1) shows the onset (at freezing temperature $T_{f}$) of the complex 
ergodicity breaking  region (the SG region). There is also an intermediate mixed phase region (SG+AF), where inside the SG, 
the magnetization of the sublattices remains finite with opposite sign. 
It should be remarked that the transition temperatures are different in both models. Particularly, for the 
same $\alpha$, $(T_{f})_{2S}>(T_{f})_{4S}$, $(T_{N})_{2S}>(T_{N})_{4S}$  and 
$ (d T_{N}/d\alpha)_{2S} < (d T_{N}/d\alpha)_{4S}$ ($T_{N}$ is the Neel temperature). These 
distinct behaviors
show the different sensitivity to the magnetic coupling between 
the two models. For the sake of completeness, in Fig. (1), it is also studied the region for $J_{0}/J<0$ 
where a ferromagnetic (FE) region is found with $m_a=m_b$. Thus, one recovers 
essentially the one lattice spin glass-ferromagnetism competition \cite{korenblit,Magal3}. 
These results show 
that, in absence of magnetic fields, 
there is a symmetry between the two sublattices.  

If $\Gamma$ is turned on (with $H=0$), two distict situations can be identified. 
For the degree of frustration $\alpha>1/\sqrt{2}$  ($0\leq J_{0}/J<\sqrt{2}$), 
the magnetization of 
both sublattices is zero. 
Therefore, the problem  is reduced to the one lattice 
problem in the presence of a 
transverse field $\Gamma$ studied in reference 
\cite{Alba2002} (see Fig. (2-a)).  
For small $\Gamma$, the 
freezing temperature $T_{f}$ is different for each model. However, when 
the temperature is decreased 
and the spin flipping increases due to $\Gamma$, the critical behavior of both models 
tends to become identical. The critical field $\Gamma_{c}$,  at the QCP for 2S and 4S models, is the same.  
The Fig. (2-b) shows a phase diagram 
for a smaller degree of frustration $\alpha=0.59$ ($J_0/J=1.7$). 
For this set of parameters, the PM solution is still found at high temperature 
for any value of $\Gamma$. 
\begin{figure}[t]
\begin{center}
	\includegraphics[angle=270,width=14cm]{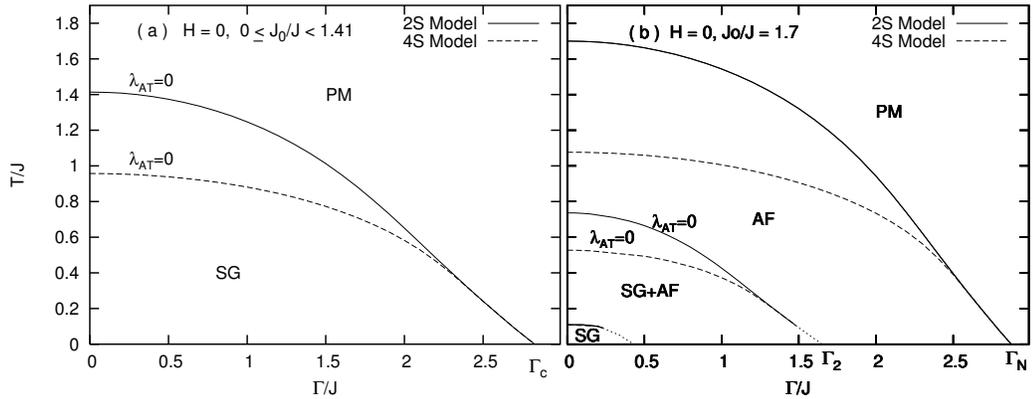}
\end{center}
\caption{Phase diagrams $T/J$ {\it versus} $\Gamma/J$ for $H=0$, (a) $0\leq
J_{0}/J< \sqrt{2}$ ($\alpha>1/\sqrt{2}$) and (b)
$J_{0}/J=1.7$ ($\alpha=0.59$).  The line conventions are the same as in Fig. (1). 
% dashed, solid, and dotted lines
% The dotted lines is the extrapolation of the curves for lower temperature.
}
\label{fig2}
\end{figure}
For small values of $\Gamma$, when the temperature is decreased, 
there is a transition to the AF phase at $(T_{N})_{2S,4S}$,  but 
$(T_{N})_{2S}>(T_{N})_{4S}$.   
For even lower temperature, there is other transition 
to the mixed phase (SG + AF) at $(T_{f})_{2S,4S}$  (again $(T_{f})_{2S}>(T_{f})_{4S}$),
and finally to the SG phase. These transitions are mainly thermally driven. On the other hand, 
when $\Gamma$ is increased, the Neel temperature $T_{N}$ for both models 
decreases towards a QCP at $\Gamma_N$. 
The temperatures $(T_{f})_{2S, 4S}$,  
given by the AT line, 
decrease when $\Gamma$ increases like the Neel temperature. 
We can observe that both 
models converge to the same critical behavior when $\Gamma$ is 
increased. 
The reasons for that behavior at $\alpha=0.59$ are the same for large degree of frustration ($\alpha>1/\sqrt{2}$).
When we see Eq. (\ref{difference}), at very low temperature, for the 4S model, the term which is integrated over $\xi_{p}$ 
is dominant and 
the distinction between the two models can be neglected.  It occurs because the contribution of 
magnetic states dominate the free energy \cite{Alba2002}. 
Therefore, there is a distinction between the models  
in the region 
where the transitions are
mostly thermally driven. 
\begin{figure}
\begin{center}
	\includegraphics[angle=270,width=14.2cm]{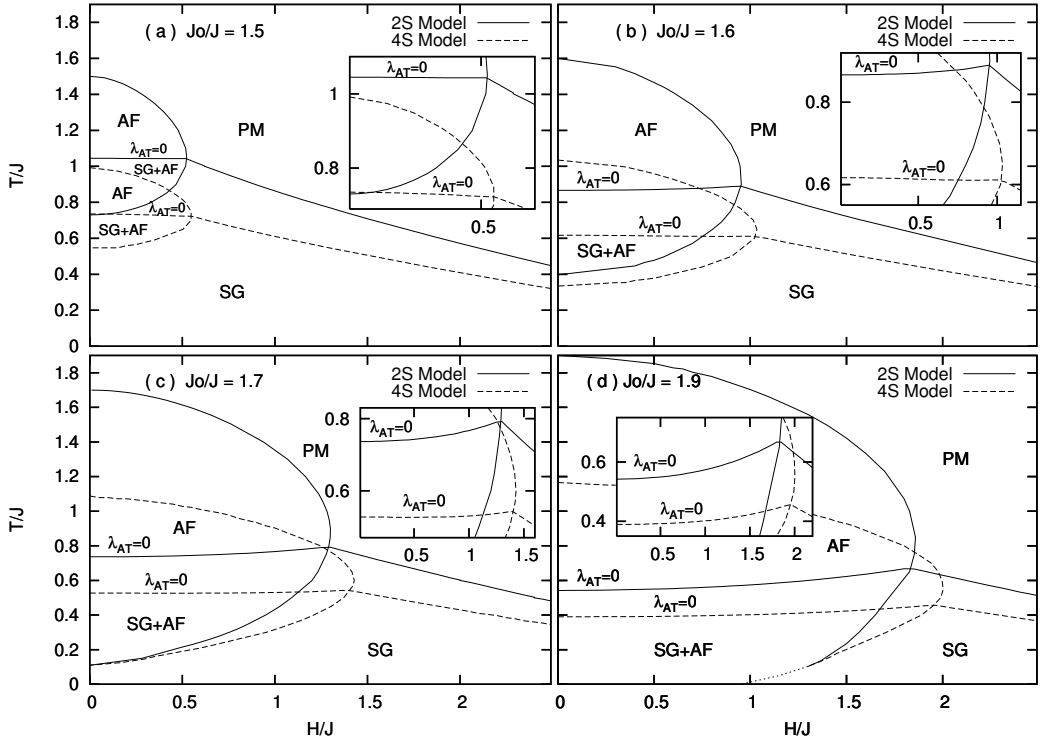}
\end{center}
\caption{Phase diagrams $T/J$ {\it versus} $H/J$ for $\Gamma=0$ and several values of $J_{0}/J$:
(a) $J_{0}/J=1.5$ ($\alpha=0.67$), (b) $J_{0}/J=1.6$ ($\alpha=0.63$), (c) $J_{0}/J=1.7$ ($\alpha=0.59$),
 and (d) $J_{0}/J=1.9$ ($\alpha=0.53$). The insets represent a zoom in
$T$ near $T_{f}$. It is used the same convention as Fig. (1) for the  line types.
}
\label{fig3}
\end{figure}

Fig. (3)  
shows the phase diagrams $T/J$ {\it versus} $H/J$ 
when $\Gamma=0$. 
These phase diagrams are studied with four different  
values of degree of frustration $\alpha$.  
The field $H$ breaks the symmetry between the sublattices and produces an effect similar to 
the KS model \cite {korenblit,korenblit2}. 
It tends to destroy the AF phase. % which is given by the order parameter $l_{m}=m_{a}-m_{b}$. 
For instance, the Neel temperatures $(T_{N})_{2S, 4S}$  
decrease fast
when $H$ is enhanced in the region $0< H < H_c$ ($H_{c}$ is the magnetic field   
when $(T_{f})_{s}=(T_{N})_{s}$  ($s=2S$, $4S$)). 
However, 
as one can see quite clearly in Figs. (3-b)-(3-d), 
there is a range of magnetic field close to $H_{c}$ 
where the freezing temperatures, associated with the AT instability, increase showing that the frustration 
is favored 
in this two-sublattice problem \cite{korenblit,korenblit2}. 
This behavior is 
different from the one lattice problem where the $T_f$ decreases monotonically \cite{SK} 
for any value of 
$H$. Nevertheless, for $H>H_{c}$, the behavior of the AT line for 2S and 4S models becomes 
similar to the one lattice problem. 
The favoring of frustration for both models as function of 
$H$ can be related to drastic decreasing of 
Neel temperature \cite{korenblit2} near $H_{c}$. 
Despite this, the presence of the magnetic field produces different effects for 
2S and 4S models. 
The  freezing temperature, given by the AT instability, increases faster
in the 2S model than in the 4S for increasing $H$. 
The AF solution is also more robust for the 4S model, it is found for greater values of $H$ 
than the 2S model.   
\begin{figure}
\begin{center}
	\includegraphics[angle=270,width=14.2cm]{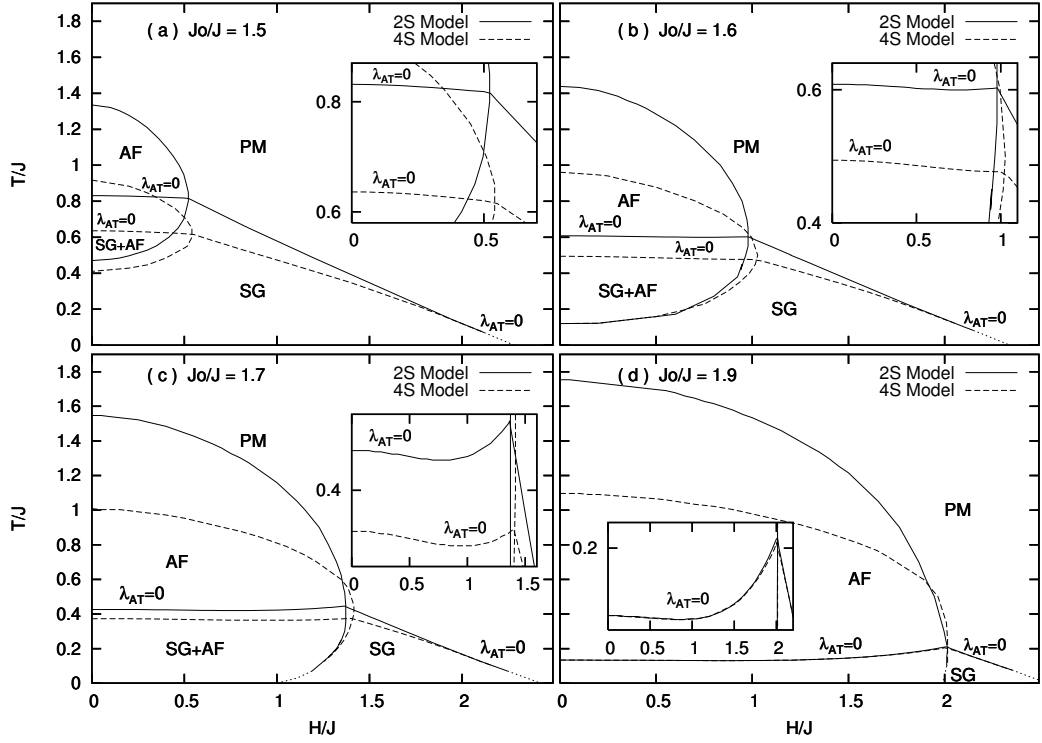}
\end{center}
\caption{Phase diagrams $T/J$ {\it versus} $H/J$ for $\Gamma/J=1.0$ and several values of $J_{0}/J$:
(a) $J_{0}/J=1.5$ ($\alpha=0.67$), (b) $J_{0}/J=1.6$ ($\alpha=0.63$), (c) $J_{0}/J=1.7$ ($\alpha=0.59$) and 
(d) $J_{0}/J=1.9$ ($\alpha=0.53$). The same 
convention as Fig. (3) is used for the line types and the insets.
}
\label{fig4}
\end{figure}

In Fig. (4), the phase diagrams $T/J$ {\it versus} $H/J$ is shown, but  with the transverse field tunned ($\Gamma/J=1.0$).  
The  set of values of $\alpha$ is the same as Fig. (3).  
For that situation, there are very important differences if compared with the $\Gamma=0$ situation.  
One effect of the applied $\Gamma$ is to depress $(T_{N})_{2S, 4S}$ and $(T_{f})_{2S,4S}$ for
any value of $H$. 
Other important effect is that the AT lines for both models have a quite 
different behavior in the range $0<H<H_{c}$. 
For instance, in Fig. (4-a) where $\alpha=0.67$ ($J_{0}/J=1.5$), 
when $H$ is enhanced until $H_{c}$, the  $T_{f}$, associated with the AT instability,
decreases faster than the equivalent phase diagram with $\Gamma=0$ 
(see Fig. (3-a)).   
When $\alpha$ is increased (see Figs. (4-b)-(4-c)), the AT line shows  the delicate balance 
between the mechanisms enforced by $\Gamma$ and $H$.
For $H\approx 0$, where the presence of $\Gamma$ is dominant, the spin flipping tends to  destroy frustration ($dT_{f}/dH<0$). 
As long as $H$ increases, the frustration is favored again. 
When $H\approx H_{c}$, it is quite clear that $dT_{f}/dH>0$. 
Particularly, close to $H_{c}$, there is a strong increase in the $T_{f}$ as $\alpha$ { decreases}.     
When the $\Gamma$ field is enhanced as in Fig. (5) (where $\Gamma/J=1.7$ and $\Gamma/J=2.4$ 
with the degree of frustration $\alpha=0.59$), 
%The degree of frustration is chosen as $\alpha=0.59$ ($J_{0}/J=1.7$)
 the SG character is entirely suppressed. 
The Neel temperature decreases towards $H_{1}$. 
The differences between 2S and 4S models still exist for $H\approx 0$. 
However, the increase of $\Gamma$ shifts the Neel temperature of both models towards zero. As consequence, 
the difference $(T_{N})_{2S}-(T_{N})_{4S}$ is increasily small.                    

\begin{figure}
\begin{center}
	\includegraphics[angle=270,width=14cm]{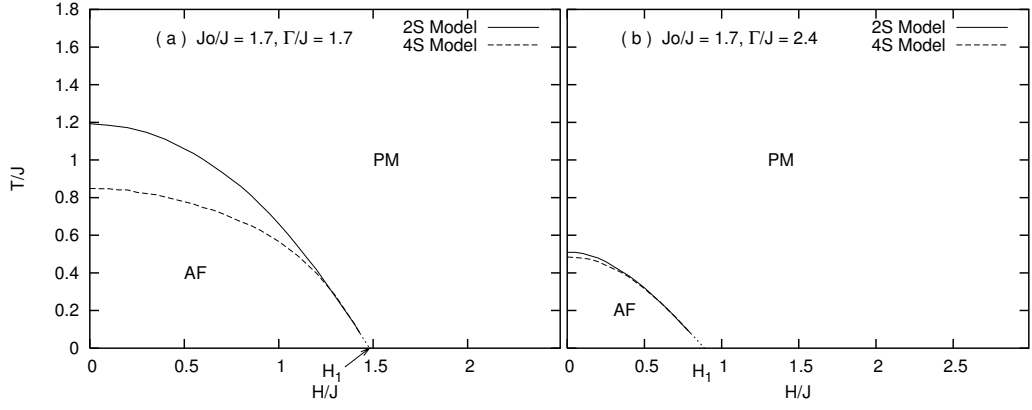}
\end{center}
\caption{Phase diagrams $T/J$ {\it versus} $H/J$ for $J_{0}/J=1.7$ ($\alpha=0.59$) and: (a)
$\Gamma/J=1.7$, (b) $\Gamma/J=2.4$. The same convention as Fig. (1) is used for line types.
}
\label{fig5}
\end{figure}
Finally, the study of the magnetic susceptibility allows one to confirm the position 
of the Neel temperatures and to study in detail the differences between the 2S and 4S models near
$T_{N}$. 
The magnetic susceptibility $\chi$ is obtained differentiating the saddle point
equations (\ref{em}, \ref{eq}, \ref{ex}) 
with respect to H. 
In the limit of zero field ($H\rightarrow 0$), the six-independent saddle point equations
system is reduced to a three-independent saddle point equations with the following
relations: $m_a(H=0)=-m_b(H=0)$, $q_a(H=0)=q_b(H=0)$ and $\bar{\chi}\equiv\bar{\chi}_a(H=0)
=\bar{\chi}_b(H=0)$. In this case, the resulting linear system for $\partial m_p/\partial H$, 
$\partial q_p/\partial H$ and $\partial \bar{\chi}_p/\partial H$ ($p=a$ or $b$) is also 
simplified giving $\chi=\partial m_a/\partial H=\partial m_b/\partial H$, 
$\partial q_a/\partial H=-\partial q_b/\partial H$, 
$\partial \bar{\chi}_a/\partial H=-\partial \bar{\chi}_b/\partial H$ \cite{korenblit}.
 Thus,
\begin{equation}
\chi=\frac{\bar{\chi}+\Pi}{T+J_o(\bar{\chi}+\Pi)}
\label{chi}
\end{equation}
where  $\bar{\chi}$ is given in Eq. (\ref{ex}) and $\Pi$ is defined in Eq. ({\ref{pi}}) of the Appendix.

% The resulting linear system for  $\partial m_p/\partial H$, $\partial q_p/\partial H$ and $\partial \bar{\chi}_p/\partial H$ is simplified taking the limit of zero field, then $\chi=\partial m_a/\partial H=\partial m_b/\partial H$, 
% $\partial q_a/\partial H=-\partial q_b/\partial H$, $\partial \bar{\chi}_a/\partial H=-\partial \bar{\chi}_b/\partial H$, 
% $m_a=-m_b$, $q_a=q_b$, and $\bar{\chi}\equiv\bar{\chi}_a=\bar{\chi}_b$ \cite{korenblit}. Thus,
%

\begin{figure}
\begin{center}
	\includegraphics[angle=270,width=14cm]{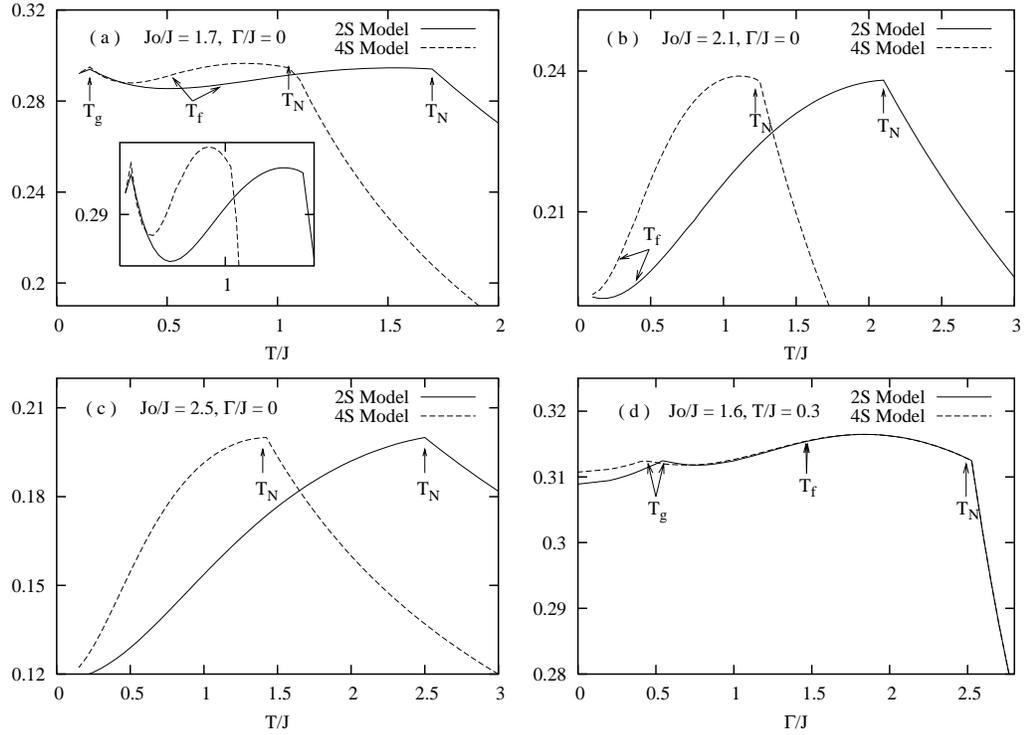}
\end{center}
\caption{The figures (a), (b) and (c) are plots of $\chi$ {\it versus} $T/J$ for 
$\Gamma=H=0$ and 
several values of $J_{0}$ (1.7; 2.1; 2.5). The figure (d) is the $\chi$ as a 
function of 
$\Gamma/J$ for $H=0, ~T=0.3J$ and $J_{0}=1.6 J$. It is used the same 
convention as Fig. (1) for the line types.
%The solid lines correspond to the 2S model and dashed lines to the 4S model. 
The inset in (a) represents a zoom of $\chi$ near the 
critical points.}
\label{figsus}
\end{figure}

For high temperature, in the PM phase, the susceptibility has the expression 
$\chi=\bar{\chi}/ (T + J_{0}\bar{\chi})$, as expected. Now, 
when the temperature is decreased, $\chi$ shows a  brusque change in its inclination
at $T_{N}$ and a cusp at $T_{g}$ ($T_g$ is the temperature below $T_f$, where 
$m_p=m_{p^{'}}=0$), but it does not show any anomaly at $T_{f}$, as we can see in 
Fig. (\ref{figsus}).  For intermediaries values of $J_{0}$, the 
susceptibility in the AF phase near $T_{N}$ does not decrease with the decrease of temperature 
(inset of Fig. (\ref{figsus}-a)). However, it exhibits a slow increase until reaching a maximum 
value, then it decreases. 
That is because of the large number of 
frustrated couplings. Nevertheless, strong antiferromagnetic average coupling, for example $J_{0}/J=2.5$, $\chi$ 
has a maximum in $T_{N}$ for both models (Fig. (\ref{figsus}-c)). In Fig. (\ref{figsus}-b), 
for $J_{0}/J=2.1$ and $\Gamma=H=0$, $\chi$ exhibits a maximum at $T_{N}$ only for the 2S model, 
while for the 4S model, $\chi$ still increases close to $T_{N}$ in the AF phase. 
 Therefore, due to the presence of non-magnetic states in the 4S model, its
frustrated couplings are less sensitive than the 2S model ones for the increasing of $J_{0}$.
%Therefore, for the same value for $\alpha$, the small number of frustrated couplings affects the 4S model with larger intensity.
Another difference of the models is in the value of the critical points at high temperatures 
as showed in Fig. (\ref{figsus}).   The Fig. (\ref{figsus}-d) 
shows the $\chi$ for $T=0.3J, ~H=0$, and $\alpha=.625$ as 
a function of $\Gamma$. Again, the critical points are marked by 
discontinuities in the $\chi$ that is obtained changing the transverse field. These discontinuities have the same 
shape described above.   

\section{Conclusion}

In the present work, the competition between antiferromagnetism (AF) 
and spin glass (SG) in disordered fermionic Ising models has been studied. The $J_{ij}$ coupling among the spins
is a random gaussian variable with average and variance given by $-4J_{0}/N$ and $32J^2/N$, respectively. There is also a magnetic field applied with components transverse 
($\Gamma$) and parallel ($H$) to the Ising spin direction. The model is similar to the Korenblit-Shender model (KS) \cite{korenblit} for classical Ising spin which consists of two sublattices where only the interlattice frustration has been considered. 

The problem has been formulated in the path integral formalism where { the} spin operators 
are represented by bilinear combinations of Grassmann fields following closely the 
approach introduced in a previous work \cite{Alba2002}, which has studied the SG critical properties in presence of a transverse field using the static approximation.  The focus in this 
work has been to investigate the boundary phase of the SG/AF competition in this two-sublattice model \cite{korenblit} for Ising spins in its fermionic representation. 
For that purpose, the static approximation has been adopted like the previous work  
in Ref. \cite{Alba2002}.  Two formulations for the fermionic representation for the 
operator $S_{i}^{z}$ have been assumed: the first one is the unrestricted four-state 
4S model (four eingenvalues, two of them magnetically insensitive); the second one is the restrict 2S model where the non-magnetic eigenvalues are forbidden by an imposed constraint. It should be remarked that the transverse component $\Gamma$ can tune spin flipping while $H$ destroys the symmetry between the sublattices. Therefore, these components have opposite roles as related to the emergence of frustration in the problem.

If $\Gamma=0$, the Fischer \cite{fischer} relation $\bar{\chi}_{p}=1-q_{p}$ ($p=a$, $b$) 
is recovered in the 2S model, as we can see from the equations (\ref{em})-(\ref{ex}).  
Thus, the results of the KS model are replicated \cite{korenblit}. 
That is an important difference with the 4S model, where the Fischer relation is not satisfied 
 due to the coupling between diagonal and off-diagonal replica matrix elements. Therefore, 
the internal field $h_{p}$ (see Eq. \ref{hp}) behaves in a quite distinct way for both models. 
These differences can be seen very clearly in the phase diagram temperature {\it versus} 
$J_{0}/J$ given in Fig. (1) with $H=0$. The onset of AF (at $T_{N}$) and the complex 
ergodicity breaking given by the Almeida-Thouless (AT) line (at $T_{f}$) 
appear for lower temperatures in the 4S model compared with the 2S one. That is a direct consequence of the distinct applied internal fields for each model.  
In the susceptibility $\chi$, given in Eq. (\ref{chi}), these differences can be seen even in 
a more detailed way (see figures (\ref{figsus}-a)-(\ref{figsus}-c)). 

When the transverse component $\Gamma$ is turned on ($H$ is kept equal to zero) (see Fig. (2)), 
there is an important change in the critical behavior for both models. 
In the range of degree of frustration $\alpha\equiv\left[J_{0}/J\right]^{-1}>1/\sqrt{2}$ 
($0\leq J_{0}/J<\sqrt{2}$), the problem is reduced to the one lattice problem studied in 
Ref. \cite{Alba2002}. For small $\Gamma$, there are two different transition temperatures 
from paramagnetism (PM) to SG for each model. These transition temperatures converge to same 
value as $\Gamma$ increases given origin to the same QCP at $\Gamma_{c}$. 
For $\alpha< 1/\sqrt{2}$, for small $\Gamma$, when the temperature is lowered, there are the 
following sequences of transitions: (1) PM/AF at $T_{N}$ ($(T_{N})_{2S}\neq (T_{N})_{4S}$);
(2) AF/(AF+SG), given by the position of the Almeida-Thouless (AT) line at 
$T_{f}$ ($(T_{f})_{2S}\neq (T_{f})_{4S}$),  where (AF + SG) is a mixed phase; 
(3) (AF+SG)/SG at $T_{g}$ ($(T_{g})_{2S}\neq (T_{g})_{4S})$, where at this transition temperature the sublattice magnetizations become null. 
For increasing $\Gamma$, the $T_{N}$ for each model decreases converging to the same transition line. 
The same effect is found for $T_{f}$. A QCP for PM/AF 
transition is found at $\Gamma_{N}/J=J_{0}/J+2J/J_{0}$.    
   
Other important consequence appears when the parallel field $H$ is turned on and, 
consequently, the symmetry between the sublattices is broken.
That is source of unusual effects because the 
internal field applied in the sublattice $p=a$, $b$ depends on the order parameters of the 
other sublattice $p'=b$, $a$ \cite{korenblit,korenblit2}.    
For the 2S model, the results of the KS model are reproduced \cite{korenblit} (see Fig. (3)). 
The Neel temperatures for both models (which are different) have a strong decrease as $H$ 
increases. 
However, the frustration is favored for a particular range of $H$ as shown in Fig. (3) from 
the AT line behavior. 
 It occurs due to the magnetization of a particular sublattice (for instance $p$ where 
$m_p>m_{p^{'}}$) which contributes to decrease the absolute value of the average internal field 
($\bar{h}_p=H-J_0m_{p^{'}}$) that acts on the other one. Hence, in this $H$ range where 
$\bar{h}_p$ decreases with $H$, $T_f$ can be increased \cite{korenblit}.
That effect is stronger when $\alpha$ is decreased ($J_{0}/J$ is enhanced), 
since the contribution of $J_0m_{p^{'}}$ on $\bar{h}_{p}$ is increased, as shown in  
Figs. (3-b)-(3-c).  Because of the presence of non-magnetic state, the 4S model has potentially a weaker response to the magnetic 
interactions. 
The results show that the AT line can also increase, but slower than the 2S model for 
the same range of $H$. Nevertheless, in the 4S model, the AF phase is more robust than 
the 2S one, remaining as solution even for higher values of $H$. 
However, for very low temperature, the contribution of the non-magnetic states in the 
4S model becomes less significant. Therefore, both models exhibit quite similar critical lines.

When the transverse field $\Gamma$ and $H$ are simultaneously turned on, the effective internal field applied in 
a particular sublattice $p$ becomes $\Delta_{p}=\sqrt{h_{p}^{2}+\Gamma^{2}}$. 
Therefore, its effects due to the presence of $H$ and $\Gamma$ start to compete. 
The effects from $H$  in the $h_{p}$           
are included explicitly, but they are also included implicitly by the order parameters. 
It should be noticed that effects of $\Gamma$ in  
the order parameters are present as well. 
As consequence, there is complex balance of the effects discussed in 
the previous paragraph, dependent on the relation $H/\Gamma$. 
Particularly, for small $H$, the contribution from $\Gamma$ is dominant and the frustration is not favored.
As long as $H$ 
enhances, there is a region in the diagram temperature {\it versus} $H$ 
where the 
frustration becomes favored (see Fig. (4)). 
Therefore, the transverse component of the magnetic field produces the same effect in each sublattice. 
It suppresses the magnetic order and the frustration. 
While, in the region where  a two-sublattice structure is characterized, the parallel 
component acts in an asymmetric way in each sublattice. In this sense, it is always against 
the AF order. Nevertheless,
it can favor frustration within a certain range of $H$.

To conclude, we studied the SG/AF competition in a two sublattice model 
\cite{korenblit} where the Ising spins have a fermionic representation.
There is a magnetic field applied. This field has a transverse component ($\Gamma$) 
which can flip the spins leading the transitions to a QCP. It has also a parallel component ($H$) which breaks the symmetry between the sublattices. These components produce opposite effects on the frustration.     
Our hope is to use this representation to study the several strongly correlated problems 
where  there is a SG/AF competition which is associated with a QCP as, for instance, in the Cerium alloy $Ce_{2}Au_{1-x}Co_{x}Si_{3}$   \cite{Majundar}. 
On the other hand, we have used the static approximation and replica symmetry ansatz. Quite recently, a new scheme for breaking replica symmetry has been proposed \cite{Kiselev} which is particularly suitable for 
the present two-sublattice problem studied here. That could be used to improve considerably the description of the SG/AF competition of our fermionic model.    
That would be an object for future work.

{\bf Acknowledgments}

The numerical calculations were, in part,  
per\-for\-med at LSC (Cur\-so de Ci\-\^en\-cia da Com\-pu\-ta\-\c{c}\~ao, UFSM) and 
at Grupo de F\'\i\-sica Estat\'\i\-tica-IFM, Universidade Federal de Pelotas.
This work was partially supported by the Brazilian agency 
CNPq (Con\-se\-lho Na\-cio\-nal de De\-sen\-vol\-vi\-men\-to 
Ci\-en\-t\'\i\-fi\-co e Tec\-no\-l\'o\-gi\-co). The authors are grateful to Prof. Alba Theumann
for useful discussions.

\appendix 
\section{Appendix}\label{apendicea}
In this appendix, the function $\Pi$ (see Eq. (\ref{chi})) is found explicitly:
%\begin{equation}
\begin{align}
\Pi&=
\frac{\beta^2J^2K_2(F_2-F_3)}{1+\beta^2J^2K_1}-
\frac{\beta^2J^2F_3}{1+\beta^2J^2F_4}\left(F_3-\frac{K_2(1+\beta^2J^2F_6)}{1+\beta^2J^2K_1}\right)~,
\\
\label{pi}
%\end{equation}
%
%\begin{align}
K_1&=F_1-\frac{2F_5(1+\beta^2J^2K_2F_6)}{1+\beta^2J^2F_4}~,\\
K_2&=F_2-\frac{2\beta^2J^2K_2F_3F_5}{1+\beta^2J^2F_4}
\end{align}
where:
\begin{align}
F_1&=\int_{-\infty}^{\infty}Dz\frac{\partial^2}{\partial h^2}\left[\frac{\int_{-\infty}^{\infty}D\xi \partial \cosh\sqrt{\Delta}/\partial h} {s+\int_{-\infty}^{\infty}D\xi\cosh\sqrt{\Delta}}\right]^2~,\\
F_2&=\int_{-\infty}^{\infty}Dz\frac{\partial}{\partial h}\left[\frac{\int_{-\infty}^{\infty}D\xi \partial \cosh\sqrt{\Delta}/\partial h} {s+\int_{-\infty}^{\infty}D\xi\cosh\sqrt{\Delta}}\right]^2~,\\
F_3&=\int_{-\infty}^{\infty}Dz\frac{\partial}{\partial h}\left[\frac{\int_{-\infty}^{\infty}D\xi \partial^2 \cosh\sqrt{\Delta}/\partial h^2} {s+\int_{-\infty}^{\infty}D\xi\cosh\sqrt{\Delta}}\right]~,\\
F_4&=\int_{-\infty}^{\infty}Dz\left[\frac{\int_{-\infty}^{\infty}D\xi \partial^4 \cosh\sqrt{\Delta}/\partial h^4}{s+\int_{-\infty}^{\infty}D\xi\cosh\sqrt{\Delta}}- \left(\frac{\int_{-\infty}^{\infty}D\xi \partial^2 \cosh\sqrt{\Delta}/\partial h^2}{s+\int_{-\infty}^{\infty}D\xi\cosh\sqrt{\Delta}}\right)^2\right]~,\\
%
%\end{align}
%\begin{align}
F_5&=\int_{-\infty}^{\infty}Dz\left[\frac{\int_{-\infty}^{\infty}D\xi \partial \cosh\sqrt{\Delta}/\partial h}{s+\int_{-\infty}^{\infty}D\xi\cosh\sqrt{\Delta}}\frac{\partial}{\partial h} \left(\frac{\int_{-\infty}^{\infty}D\xi \partial^2 \cosh\sqrt{\Delta}/\partial h^2}{s+\int_{-\infty}^{\infty}D\xi\cosh\sqrt{\Delta}}\right)\right]~,\\
F_6&=\int_{-\infty}^{\infty}Dz\frac{\partial^2}{\partial h^2}\left[\frac{\int_{-\infty}^{\infty}D\xi \partial^2 \cosh\sqrt{\Delta}/\partial h^2} {s+\int_{-\infty}^{\infty}D\xi\cosh\sqrt{\Delta}}\right]~,
\end{align}
with $\Delta=h^2+\beta^2\Gamma^2$ and $h\equiv \beta h_{a}=
\beta h_{b}$ ($h_a$ is defined in Eq. (\ref{hp})).

\end{document}